\documentclass[prb,twocolumn,showpacs,preprintnumbers,amsmath,amssymb]{revtex4}
\usepackage{graphicx}
\usepackage{dcolumn}
\usepackage{bm}
\begin{document}

\title{Anisotropic spin-fluctuations in SmCoPO revealed by $^{31}$P NMR measurement}
\author{Mayukh Majumder$^1$, K. Ghoshray$^1$\thanks{E-mail address: kajal.ghoshray@saha.ac.in}, A. Ghoshray$^1$, Anand Pal$^2$, and V.P.S. Awana$^2$}
\affiliation{$^1$ECMP Division, Saha Institute of Nuclear Physics\\\address{1/AF Bidhannagar,
Kolkata-700064, India}\\
$^2$Quantum Phenomenon and Applications Division, National Physical Laboratory (CSIR)\\
\address{New Delhi-110012, India}}

\begin{abstract}
$^{31}$P NMR spectral features in polycrystalline SmCoPO reveal an axially symmetric local magnetic field.
 At low temperature, the anisotropy of the internal magnetic field increases rapidly, with $K_{ab}$
 increasing faster than that of $K_c$. The dominant contribution to this anisotropy arises from Sm-4$f$ electron contribution over that of Co-3$d$. The intrinsic width 2$\beta$ deviates from linearity with respect to bulk susceptibility below 170 K due to the enhancement of ($1/T_2)_{dynamic}$, which along with the continuous increase of anisotropy in the internal magnetic field is responsible for the wipe out effect of the NMR signal, well above $T_C$. $1/T_1$ shows large anisotropy confirming a significant contribution  of Sm-4$f$ electron spin fluctuations to $1/T_1$, arising from indirect RKKY type exchange interaction indicating a non-negligible hybridization between Sm-4$f$ orbitals and the conduction band, over the itinerant character of the Co-3$d$ spins. This anisotropy in originates from the orientation dependence of $\chi^{\prime\prime}(\textbf{q}, \omega$). The 3$d$-spin fluctuations in the $ab$ plane is 2D FM in nature, while along the $c$-axis, a signature of a weak 2D AFM spin fluctuations superimposed on weak FM spin-fluctuations even in a field of 7 T and far above $T_N$ is observed. The enhancement of this AFM fluctuations of the Co-3$d$ spins along $c$-axis, at further low temperature is responsible to drive the system to an AFM ordered state.
 \end{abstract}

\maketitle
\section{INTRODUCTION}

The unconventional
nature of the iron based (grouped in several families)
superconductors has drawn immense attention from the theoreticians
as well as experimentalists.\cite{Johnston10} Presence of strongly correlated electrons
are responsible for diverse electronic and magnetic properties shown
by these materials. The non-superconducting parent compounds also
show interesting properties\cite{Mcguire08,Chen08} such as spin
density wave (SDW) transition, structural phase transition,
itinerant ferromagnetism etc. Several members of these families show
superconductivity (SC) upon carrier doping. In 1111 and 122 family,
superconductivity can be achieved by Co doping in place of
iron.\cite{sefat08,sefat2-08} It is presumed that the study of Co
based non superconducting members may provide useful
information about the key factor that determines the ground state
i.e. either SC or magnetic.

The magnetic property of the \emph{RE}CoAsO (\emph{RE} = rare earth) series has been investigated \cite {sefat08,Yanagi08,Ohta09,Ohta209,Marcinkova10,McGuire10,Awana10,Ohta10,Ohtajpn10,APal11} for both non-magnetic (La), and magnetic (Ce, Pr, Sm, Nd, and Gd) members, the \emph{RE}CoPO on the other hand has been reported for LaCoPO\cite{Yanagi08,Majumder09,Sugawara09,Majumder10} and CeCoPO.\cite{Krellner09} Both these compounds exhibit only ferromagnetic transition due to Co-3d electrons with $T_C$ = 35 K and 75 K respectively. The Ce-ions are on the border to magnetism with a Kondo scale of $T_K$$\sim$40 K with enhanced Sommerfield coefficient of $\gamma$=200mJ/molK$^2$.
In \emph{RE}CoAsO series, La, Ce, and Pr show paramagnetic (PM) to ferromagnetic (FM) transition,\cite{Ohta209} whereas Sm, Nd,
and Gd show PM $\rightarrow$ FM $\rightarrow$ antiferromagnetic (AFM) transition.\cite{Ohta209,Awana10} The AFM transition was proposed to be mediated by the interaction between the \emph{RE}-4$f$ and the Co-3$d$ electrons. Furthermore, in SmCoAsO and
NdCoAsO a second AFM transition only due to \emph{RE} ion was also reported.\cite{McGuire10,APal11} Recently it has been shown from
magnetization and specific heat measurements that Sm/NdCoPO also undergo three magnetic transitions i.e. $T_{C,Co}$ (80 K), the Sm$^{4f}$-Co$^{3d}$/Nd$^{4f}$-Co$^{3d}$ interplayed AFM transition ($T_{N1}$) below 20 K and finally Sm$^{3+}$/Nd$^{3+}$ spins individual AFM transitions at ($T_{N2}$)=5.4/2.0 K.\cite{Pal11} The important difference between these two series is that in \emph{RE}CoAsO, the $T_C$ increases from La to Ce  and remains unchanged for Pr - Gd, whereas in \emph{L}CoPO family, the $T_C$ increases progressively as we go down the series from La to Sm. Thus the strength of the exchange interactions changes as one replaces As by P. In both the series, the lattice volume decreases across the series. In general, with the application
of chemical or physical pressure, $T_C$ decreases due to the increment of density of state (DOS) at Fermi level (magneto-volume
effect). However, due to the lattice size decrement, the three dimensionality of the magnetic interaction may enhance causing an
increment of $T_C$, thereby confirming the active role of the competing phenomena governing the actual ground state.

Our earlier $^{31}$P and $^{139}$La NMR measurements in grain aligned $(c\|H_0)$ LaCoPO (quasi 2D Fermi surface) reveal that the spin fluctuation of 3$d$
electrons in PM state is basically two dimensional (2D) in nature with non negligible 3D part and it is 3D in the FM
state.\cite{Majumder09, Majumder10} Moreover, relaxation rate shows weak anisotropy. Since SmCoPO has the minimum unit cell volume in this series, it would be interesting to study the paramagnetic state to probe the interplay between increasing interlayer interaction due to three dimensionality of the Fermi surface (causes increment of $T_C$) and magneto-volume effect (causes decrement of $T_C$). Probing dynamic spin susceptibility, the
spin-lattice relaxation rate provides microscopic information on the dimensionality of spin-fluctuations. We would thus examine a few pertinent questions: (i) is the decrement of lattice volume in SmCoPO low enough to make the spin-fluctuation 3D in nature even in the paramagnetic state? (ii) Whether any anisotropy is expected in the nuclear relaxation rate in SmCoPO because the contributions of different 3$d$ orbitals of Co and 4$f$ orbitals of Sm to the Fermi surface, governing nuclear relaxation, would change as a result of the lattice shrinkage? and (iii) to understand the mechanism which drives the FM oriented Co-3$d$ spins to reorder antiferromagnetically at further low temperature which persists even in a field of 14 T.

\section{EXPERIMENTAL}
 Polycrystalline samples of SmCoPO and LaCoPO were synthesized by solid state reaction the details of which are described in \cite{Pal11}.  The powder sample was characterized using x-ray diffraction technique with CuK$\alpha$ radiation at room temperature in a Rigaku X-ray diffractometer. The Rietveld analysis of the X-ray powder diffraction data confirmed that the samples are crystallized in tetragonal phase with all the peaks indexed to the space group P4/nmm. The $^{31}$P NMR measurements were carried out in powder samples of SmCoPO and LaCoPO, using a conventional phase-coherent
 spectrometer (Thamway PROT 4103MR) with a 7.0 T($H_0$) superconducting magnet (Bruker). The temperature variation study was performed in an Oxford continuous flow cryostat equipped with a ITC503 controller. The spectrum was recorded by changing the frequency step by step and recording the spin echo intensity by applying a $\pi/2- \tau - \pi/2$ solid echo pulse sequence. Shifts were measured with respect to the $^{31}$P resonance line position ($\nu_R$) in H$_3$PO$_4$ solution. The spin-lattice relaxation time ($T_1$) was measured using the saturation recovery method, applying a single $\pi$/2 pulse. The spin-spin relaxation time ($T_2$) was measured applying $\pi/2 -\tau - \pi$ pulse sequence.

\section{RESULTS AND DISCUSSIONS}

\subsection{$^{31}$P NMR spectra in SmCoPO }
The Hamiltonian for the interaction between the nuclear and
electronic spins in the presence of external field H$_0$ can be
written as
\begin {equation}
H = -\gamma \hbar \textbf{I}.\textbf{H}_0+ \sum_j \textbf{I}. A_j
.\textbf{S}_j + \sum_j \gamma \hbar \textbf{I}.\mu_B g
\textbf{S}_j\frac{3\cos^2 \theta_j-1}{r^3_j},
\end {equation}
where the first term is the nuclear Zeeman energy, the second term
represents hyperfine interaction with jth magnetic ion having spin
S$_j$ and the third term denotes the dipolar interaction. In the
most general case, when a nucleus experiences a completely
anisotropic internal magnetic field (sum of the hyperfine and
dipolar field), the resonance frequency in a single crystal is given
by\cite{Bloembergen53}
\begin {equation}
\nu = \nu_R[1+K_{iso}+K_{ax} (3\cos^2\theta-1)+K_{aniso} \sin^2\theta
\cos^2 2\phi],
\end {equation}
where $K_{iso} = (K_1+K_2+K_3)/3$, $K_{ax} =
(2K_3-K_1-K_2)/6$, and $K_{aniso} = (K_2-K_1)/2$. $K_1$, $K_2$, and $K_3$ are the principal
components of the total shift tensor. If a
nucleus experiences an internal field of cylindrical symmetry, the
third term in eq. (2) vanishes, since $K_1 \approx K_2$.
\begin{figure}
{\centering {\includegraphics{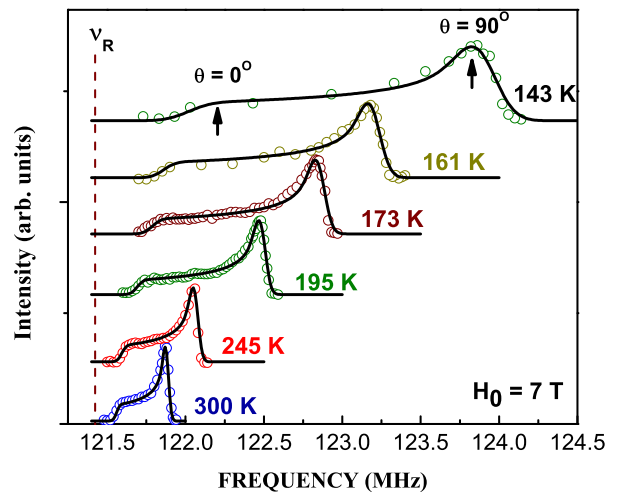}}\par} \caption{Typical $^{31}$P NMR spectrum ($\circ$) in SmCoPO. Vertical dashed line
represents reference position. Continuous lines
correspond to the simulated spectrum derived from eq. 3 using Gaussian line shape. The vertical arrows indicate step $(\theta = 0^\circ)$ and maximum $(\theta = 90^\circ)$ respectively.} \label{spectra}
\end{figure}

In a polycrystalline specimen, the crystallites being oriented
randomly, the anisotropic shift results in a broadening
proportional to the applied field. Since all values of
$u=\cos\theta$ are equally probable the expression for the line shape would be $p(\nu)\sim$
1/$\mid d\nu/du\mid$. Superimposing a gaussian broadening of width
2$\beta$, to the resonance line from each of the crystallites, the
line shape in polycrystalline sample will be
\begin {equation}
I(\nu') = \int^\infty_{-\infty}
p(\nu)\exp[-(\nu-\nu')^2/2\beta^2]d\nu.
\end {equation}
The shift parameters $K_{iso}$,
$K_{ax}$ and linewidth (2$\beta$) can be obtained by fitting the spectra using eq. (3).

 Figure 1 shows some typical $^{31}$P NMR spectra in
polycrystalline SmCoPO at different temperatures. The resonance line
shape corresponds to a powder pattern for a spin $I = 1/2$ nucleus
experiencing an axially symmetric local magnetic field, as expected
for SmCoPO having tetragonal symmetry. The step in the low-frequency
side corresponds to $\textrm{H}_0\|c$ $(\theta = 0^\circ)$ and the
maximum in high frequency side corresponds to $\textrm{H}_0\perp c$ $(\theta = 90^\circ)$. The shift of the step with respect to
the reference position ($\nu_R$), corresponds to $K_c$ and that of
the maximum corresponds to $K_{ab}$, where $K_{iso} = \frac{2}{3}K_{ab} +
\frac{1}{3}K_{c}$ and $K_{ax} = \frac{1}{3}(K_{c}-K_{ab})$. The continuous line
superimposed on each experimental line is the calculated spectrum
corresponding to eq. 3 using Gaussian line shape. Most important feature is that the separation between the step, $K_c$ and the maximum, $K_{ab}$ increases at low temperature along with line broadening. In particular, $K_{ab}$ shows much larger change than that of $K_c$. Finally, the resonance line could not be detected below 130 K, ($T_C$ at H=7 T is about 110 K as determined from the derivative
of the $\chi$ versus $T$ curve; not shown here). The line did not reappear till the lowest temperature. Such thing did not happen in case of $^{31}$P and $^{139}$La NMR studies in LaCoPO, where the resonance line was detected\cite{Majumder10} even below $T_C$.

\subsubsection{$^{31}$P NMR wipe out and spin-spin relaxation rate $1/T_2$}
Figure 2 shows the variation of 2$\beta$ with the bulk magnetic susceptibility, $\chi$ (with temperature as implicit parameter)
 in SmCoPO, depicting a linear $\chi$ dependence of 2$\beta$ in the range 173-300 K. This arises mainly from the contribution due to the demagnetizing field.
Below this range there is a significant deviation from linearity,
showing a large enhancement. To get a more quantitative picture,
$^{31}$P spin-spin relaxation time ($T_2$) also known as the
transverse relaxation time was measured, as a function of
temperature (inset (b) of Fig. 2). In this measurement, the echo
integral (which arises from the transverse magnetization) was taken
as a function of time delays ($\tau$) between two $rf$ pulses. The recovery
of the transverse magnetization was found to be exponential at all
temperatures. $T_2$ was obtained by fitting the equation $M(2\tau) =
M_0\exp(-2\tau/T_2)$, (solid lines in the inset (a) in Figure 2.)
where $M_0$ is the
initial magnetization. At any temperature, the magnitude of $T_2$
remains same when measured at the position of the maximum
($\theta=90^\circ$) and at the step ($\theta=0^\circ$) respectively,
indicating its isotropic nature.

In general, 1/T$_2$ can be written as the sum of the contribution
from (i) dipolar interaction between the nuclear magnetic moments
which is temperature and field independent (1/T$_2|_{static}$) and
(ii) the dipolar and hyperfine interactions of the nuclei with the longitudinal component of the
fluctuating magnetic field produced by the neighboring Co$^{2+}$
3$d$-spins and Sm 4$f$-spins. This dynamic part 1/T$_2|_{dynamic}$, is temperature
dependent, when the fluctuation frequency becomes close to nuclear resonance frequency below a certain temperature, due to the development of short range correlation among the electronic spins as $T_C$ is approached. In weak collision fast motion approximation, one can
express 1/T$_2|_{dynamic}$, in terms of the spectral density of the
spin-fluctuating hyperfine field at zero frequency,\cite{Slichter,Belesi07}
\begin {equation}
1/T_2|_{dynamic} = {\gamma_n}^2 <{\delta H_z}^2> \tau(T) + 1/2T_1
\end {equation}
where $\delta H_z$ is the local longitudinal field originating from
a magnetic moment sitting at a distance $r$ apart from the probed
nuclear spin, and $\tau$ stands for correlation time, which is only
determined by the dynamics of the exchanged coupled magnetic ions. Therefore, the observed $\theta$ independent behavior of $T_2$, as mentioned above indicates that the longitudinal component of the fluctuating electronic magnetic field at the $^{31}$P nuclear site is isotropic in nature. It is to be noted that the intrinsic width, 2$\beta$ in Eq. 3 contains sum of the contributions from 1/T$_2|_{static}$, 1/T$_2|_{dynamic}$ and that due to the demagnetizing field. Among them the first contribution is magnetic field independent while the second and third depend on field and the magnetic susceptibility.\cite{Lomer62}

\begin{figure}
{\centering {\includegraphics{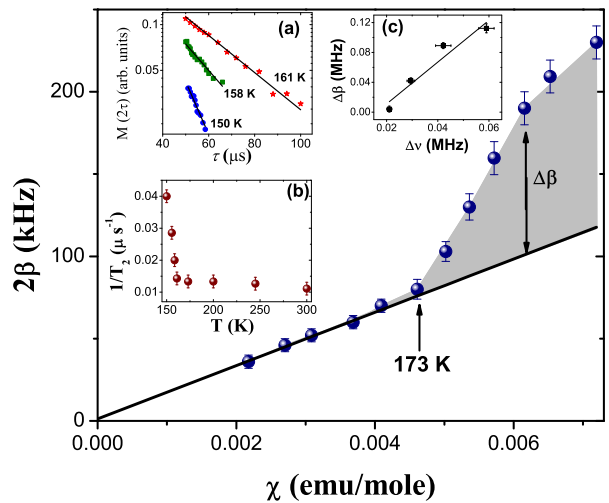}}\par} \caption{The variation of 2$\beta$ against magnetic susceptibility
(solid circle) in SmCoPO, solid line is the linear fit. Inset (a):
recovery of transverse magnetization at different temperatures, solid lines corresponds to $M(2\tau) = M_0\exp(-2\tau/T_2)$; (b)
1/T$_2$($\mu$s$^{-1}$) versus temperature; (c): $\Delta\nu$ (MHz)
versus $\Delta\beta$ (MHz), solid line is the linear fit as discussed in the text. }
\label{structure}
\end{figure}

The contribution 2$\beta_{dynamic}$ to the total 2$\beta$ was estimated, by subtracting the contribution due to the linear part determined from the extrapolated values (larger $\chi_M$ values in Fig. 2) from 2$\beta$. In the inset (c) of Fig. 2, we have plotted this 2$\beta_{dynamic}$ denoted as $\Delta\beta$ versus the line width ($\Delta\nu$) obtained from the measured $T_2$. The linear behavior confirms that the observed large enhancement of 2$\beta$ below 173 K arises due to the enhancement of 1/$T_2$ or more specifically (1/$T_2$)$_{dynamic}$. As $T_2$ reaches a value of 25 $\mu$s at 150 K, there is a possibility that at lower temperature, it becomes so short that one has to apply a delay time $\tau$ between the two $rf$ pulses ($\pi/2- \tau - \pi/2$) used to observe the spin echo, which is comparable to or shorter than the dead time of the spectrometer. As a consequence, the NMR signal coming from the $^{31}$P nuclei can not be digitized by the spectrometer, resulting the whole signal to vanish.

\begin{figure}
{\centering {\includegraphics{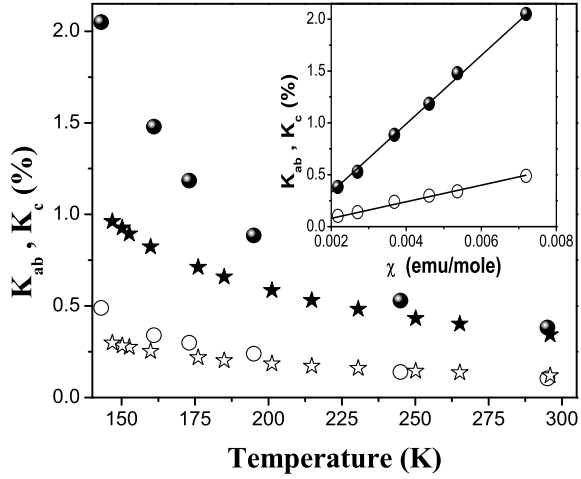}}\par} \caption{$K_{ab}$, $K_c$ vs temperature for SmCoPO (filled
circle and open circle respectively) and $K_{ab}$, K$_c$ vs
temperature for LaCoPO (filled star and open star respectively),
Inset shows $K_{ab}$, $K_c$  vs $\chi$ for SmCoPO and the solid
line is the linear fit.} \label{structure}
\end{figure}
\begin{figure}
{\centering {\includegraphics{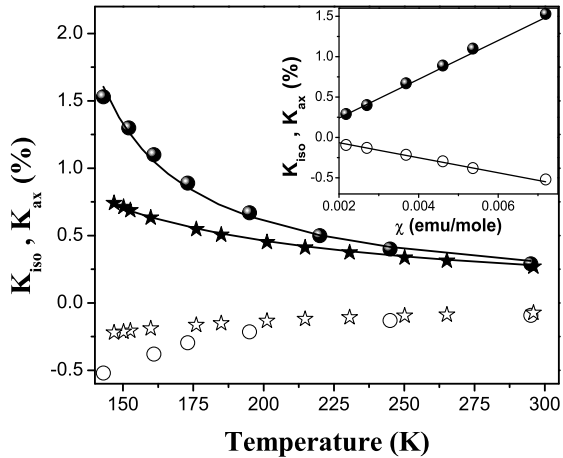}}\par} \caption{$K_{iso}$,
$K_{ax}$ vs temperature for SmCoPO (filled circle and open circle
respectively) and $K_{iso}$, $K_{ax}$ vs temperature for LaCoPO (filled
star and open star respectively), solid lines correspond to Eq. 6.
Inset shows $K_{iso}$, $K_{ax}$ vs $\chi$ for SmCoPO and the solid
line is the linear fit.} \label{structure}
\end{figure}

\subsection{Knight shift and hyperfine field}
 Figures 3 and 4 show temperature dependence of shift parameters $K_c$, $K_{ab}$, $K_{iso}$ and
$K_{ax}$ in SmCoPO and the same
in LaCoPO for comparison. Around 300 K, shift parameters for LaCoPO and SmCoPO are of same magnitude, however, at low
temperature, $K_{ab}$, $K_c$, $K_{iso}$ and $K_{ax}$ for SmCoPO
increases rapidly than in LaCoPO. Measured shift can be written as
$K = K_0 + K(T)$, where $K_0$ is the temperature independent
contribution arising from conduction electron spin susceptibility,
orbital susceptibility and diamagnetic susceptibility of core
electrons. $K(T)$ arises from the temperature dependent
susceptibility due to Co-3$d$ and Sm-4$f$ spins,
\begin {equation}
K(T) = (H_{hf}/N\mu_B)\chi(T)
\end {equation}
$H_{hf}$ is
the total coupling constant due to the electron nuclear hyperfine and dipolar interactions, $N$ is the avogadro number and $\mu_B$ is the
Bohr magneton. Insets of Figs. 3 and 4 show linear variation of $K_c$, $K_{ab}$,
$K_{iso}$ and K$_{ax}$ with $\chi = M/H$. From these plots the obtained values of the coupling constants are
$H_{hf}^{ab}$, $H_{hf}^c$, $H_{hf}^{iso}$, $H_{hf}^{ax}$ are 18.4, -4.46, 13.4 and -5.19 kOe/$\mu_B$ respectively. Using the atomic coordinates of Sm, Co and P, $H_{dip}^{ax}$ (-0.44 kOe/$\mu_B$) at the $^{31}$P site was calculated from Eq. 9 of sec.III.C. The value is one order of magnitude smaller than that of experimental $H_{hf}^{ax}$. Thus the observed temperature dependent anisotropic part of the shift is mainly due to the hyperfine interaction. If we consider that the contribution to $K_{ax}$ due to Co-3$d$ electrons is nearly same in LaCoPO and SmCoPO, then the observed larger enhancement of $K_{ax}$ in SmCoPO at low temperature compared to that in LaCoPO ( Fig. 4) is a signature of more pronounced contribution of Sm-4$f$ electrons  over that of Co-3$d$ for producing anisotropic local magnetic field at the $^{31}$P site.

Temperature dependence of $K_{iso}$ can be well described by the Curie-Weiss type behavior,
\begin {equation}
K_{iso}(T) = (H_{hf}^{iso}/N\mu_B)\frac{C}{(T-\theta)}
\end {equation}
as represented by the continuous line in figure 4 for LaCoPO and SmCoPO. The estimated P$_{eff}$ value from Curie-Weiss constant (C) are 1.4$\mu_B$ with $\theta$=53 K for LaCoPO and 1.65$\mu_B$ with $\theta$=110 K for SmCoPO, which are in close agreement with those determined from magnetic susceptibility data.\cite{Pal11} This indicates that their is a contribution of Sm 4$f$ moment over that of Co 3$d$ moment even in the paramagnetic state. It is to be noted that in NdFeAsO$_{1-x}$F$_x$ and CeCoAsO the $^{75}$As Knight shift\cite{Jeglic09,Sarkar10} was found to be
influenced by 4$f$ moments though As is situated in a different plane. A notable difference between FeAs based systems and CoP/CoAs based systems is that the hyperfine field at the $^{75}$As site in LaFeAsO$_{(1-x)}$F$_x$ is temperature independent and the temperature dependence appears only when La is replaced by other rare earths.\cite{Jeglic09,Prando10} Whereas, in CoP/CoAs based systems, the hyperfine field is temperature dependent even in LaCoPO\cite{Majumder09,Sugawara09} and LaCoAsO\cite{Ohta10}. Substitution of other rare earth gives an additional temperature dependent contribution to the shift arising from 4$f$ electrons superimposed on  that due to Co 3$d$ electrons.
\subsection{ Nuclear spin-lattice relaxation rate $1/T_1$ }

$1/T_1$ was determined from the recovery of the longitudinal component of the nuclear magnetization M($\tau$) as a function of the delay time $\tau$ using equation
\begin {equation}
M(\tau) = M(\infty)(1-\exp^{-\tau/T_1})
\end {equation}
for nuclear spin $I$=1/2. The recovery curves (inset (a) of Fig. 5) were found to be exponential throughout the whole temperature range, as expected for an ensemble of $I$=1/2 nuclei with a common spin temperature. This confirms good sample homogeneity with negligible amount of phosphorous containing impurity phase. The temperature dependence of the $^{31}$P nuclear spin-lattice
relaxation rates $(1/T_1)_{ab}$ and $(1/T_1)_c$ in SmCoPO (inset(b) of Fig. 5) in the temperature
range 140 - 300 K clearly show the anisotropic nature. The $T_1$ values in SmCoPO are
one order of magnitude shorter than that in LaCoPO near 300 K and becomes two orders of magnitude shorter near 140 K.  In case of SmFeAsO$_{1-x}$F$_x$ $^{19}$F NMR\cite{Prando10}
relaxation rate ($1/T_1$) was three orders of magnitude higher than that in
LaFeAsO$_{1-x}$F$_x$.\cite{Ahilan08} This enhancement in 1/$T_1$ was
also observed in case of $^{75}$As $1/T_1$ in 1111 superconductor
with Pr and Nd as rare earth element.\cite{Yamashita10}
The enhancement in $1/T_1$ was assigned due to the 4$f$
electrons and not due to the Fe 3$d$ electrons. Prando
el al.\cite{Prando10} have concluded that the increment of $1/T_1$
from 200 K in SmFeAsO$_{1-x}$F$_x$ was due to the Sm 4$f$ electrons
via the indirect RKKY exchange coupling and which indicates that
their is a non-negligible hybridization between Sm 4$f$ with
conduction electrons i.e. the 4$f$ electrons are not fully localized in
nature. In SmCoPO the large enhancement in magnitude of 1/T$_1$ at low temperature with
respect to that in LaCoPO should also arise due to the contribution of Sm 4$f$ electrons through the magnetic dipolar interaction and the hyperfine interaction (through RKKY type conduction electron mediated $c-f$ exchange) over the
contribution of Co 3$d$ electrons (which is also present in LaCoPO). Moreover, the continuous enhancement of
$1/T_1$ in SmCoPO from below 300 K, indicates the signature of the development of short range
correlation far above $T_C$. This should be characteristic of itinerant
magnetism of both Co 3$d$ and Sm 4$f$ electrons.\cite{Moriya85} So
$^{31}$P will feel the effect of itinerant Co 3$d$ via the
hybridization between Co 3$d$ and P 2$p$ and also feel the effect of
Sm 4$f$ electrons via RKKY interaction through the conduction
electrons along with the magnetic dipolar interaction with the same .
\begin{figure}
{\centering {\includegraphics{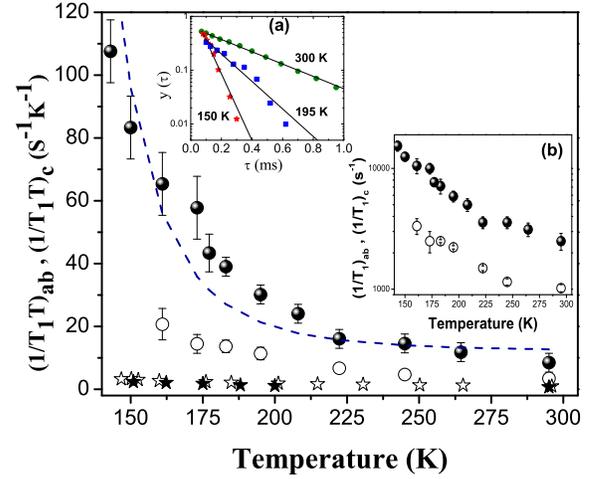}}\par}
\caption{$(1/T_1T)_{ab}$ (filled circle),$(1/T_1T)_c$ (open circle)
versus T for SmCoPO and $(1/T_1T)_{ab}$ (filled star), $(1/T_1T)_c$
(open star) versus T for LaCoPO. The dashed line corresponds to Eq.
16. Inset(a): recovery curves at different temperatures where y($\tau$) = $\frac{M(\infty)-M(\tau)}{M(\infty)}$ and Inset
(b): $(1/T_1)_{ab}$ (filled circle)and $(1/T_1)_c$ (open circle) vs
T for SmCoPO.}
\end{figure}

Contribution to spin-lattice relaxation rate due to 4$f$ spin fluctuation (from Sm) via the dipolar coupling\cite{Moriya56} is
\begin {eqnarray}
(1/T_1)_\textrm{dip} = &&\frac{\sqrt{2\pi}\gamma_n^2(g\mu_B)^2}{6\omega_e}J(J+1)\sum_i r_i^{-6}\nonumber\\
 &&\times[F_i(\alpha_i,\beta_i,\gamma_i) + F^\prime_i(\alpha_i,\beta_i,\gamma_i)],
\end {eqnarray}
where $F_i(\alpha_i,\beta_i,\gamma_i)$ and $F^\prime_i(\alpha_i,\beta_i,\gamma_i)$ are geometrical factors which depend on $\alpha_i$, $\beta_i$, $\gamma_i$, the direction cosines of $r_i$ connecting the $i$-th Sm atom and $^{31}$P nuclear spin with respect to the principal axes of the dipolar field tensor. Using the structural parameters\cite{Pal11} for SmCoPO we have calculated the dipolar field at the $^{31}$P nuclear site arising from Sm moment with the formula,
\begin{equation}
H_\textrm{dip} = \mu\sum\frac{(3r_jr_k - r^2\delta_{jk})}{r^5}; j,k=x, y, z
\end{equation}
$\mu$ is the magnetic moment of Sm moment.
The direction of the principal components of the dipolar tensor coincide with the crystallographic axes system and this makes calculation of $F_i(\alpha_i,\beta_i,\gamma_i)$ and $F^\prime_i(\alpha_i,\beta_i,\gamma_i)$ simple. Thus the lattice sum of eq. 8
\begin{equation}
\sum_i r_i^{-6} [F_i(\alpha_i,\beta_i,\gamma_i) + F^\prime_i(\alpha_i,\beta_i,\gamma_i)]= 2.143\times 10^{46} \textrm{cm}^{-6}.\nonumber
\end{equation}
The electronic exchange frequency $\omega_e$ is estimated from the Neel temperature ($T_N$= 5.4 K) of the Sm-4$f$ moment\cite{Pal11} by
\begin{equation}
(\hbar\omega_e)^2=\frac{1}{6z}\frac{(3k_BT_N)^2}{J(J+1)}.
\end{equation}
Taking $z$ =4, number of the nearest neighbor Sm spins $\omega_e=1.46\times10^{11}$ sec$^{-1}$. Finally, $(1/T_1)_\textrm{dip}$= 434 sec$^{-1}$.
At 300 K the average value of 1/$T_1$ in SmCoPO for $\theta$=0 and $\theta$=$\pi$/2 is $\sim$ 1760 sec$^{-1}$. The value of 1/$T_1$ at 300 K in LaCoPO is $\sim$ 300 sec$^{-1}$. If we now assume that the contribution due to Co-3$d$ spins is nearly same in LaCoPO and SmCoPO, then also [(1/T$_1$)$_{SmCoPO}$-(1/T$_1$)$_{LaCoPO}$] is more than three times larger than $(1/T_1)_\textrm{dip}$ for Sm 4$f$ spins i.e. the main contribution to 1/$T_1$ in SmCoPO comes from hyperfine interaction with the Sm-4$f$ spins.

Main panel of Fig. 5 shows the $(1/T_1T)_{ab}$ and $(1/T_1T)_c$ vs
$T$ curves in SmCoPO and in LaCoPO. In SmCoPO $(1/T_1T)_{ab}$
increases much faster than that of $(1/T_1T)_c$. Whereas, in LaCoPO
this anisotropy is negligible. This enhanced anisotropy is a
signature of contribution of anisotropic Sm 4$f$ orbitals along with
the Co 3$d$ orbitals. In general, ($1/T_1T)_{SF}$ is given by
\begin {equation}
(1/T_1T)_{SF} \propto \sum_q |H_{hf}(q)|^2\chi\prime\prime(q, \omega_n)/\omega_n
\end {equation}
where $\chi^{\prime\prime} (q, \omega_n$) is the imaginary part of
the transverse dynamical electron spin susceptibility, $\gamma_n$
and $\omega_n$ are the nuclear gyromagnetic ratio and Larmor
frequency respectively. $H_{hf}(q)$ is the hyperfine form factor. Terasaki\cite{Terasaki09} et al. in LaFeAsO$_{0.7}$ and
Kitagawa\cite{Kitagawa08} et. al. in BaFe$_2$As$_2$ had shown that at P site $H_{hf}(q)$ is
non-zero for $q = 0$ and also non-zero for $q\neq$0 (inplane
off-diagonal pseudo-dipolar hyperfine field has a non zero value
along c axis for $q$ = ($\pm\pi$, $\pm\pi$)). $q$$\neq$0
contribution have been found at As site for
BaFe$_2$As$_2$,\cite{Kitagawa08} SrFe$_2$As$_2$,\cite{Kitagawa09}
LaFeAsO, LaFeAsO$_{1-x}$F$_x$\cite{Ishida10} and also in Co doped
BaFe$_2$As$_2$\cite{Ning09,Ning10,Ishida10} systems. This indicates
that $1/T_1T$ at P site can feel ferromagnetic spin-fluctuations as
well as antiferromagnetic spin-fluctuations.

The directional dependence in $(1/T_1T)_{ab}$ and $(1/T_1T)_c$
can be due to anisotropy in $H_{hf}$ or due to anisotropy of $\chi\prime\prime(q, \omega_n$),
or both. Inset of Fig. 6 shows that the ratio
($1/T_1T)_{ab}$/$(1/T_1T)_c$ is higher than that of the
ratio [$(H_{hf}^{ab}$)$^2$ + $(H_{hf}^c$)$^2$]/2$(H_{hf}^{ab}$)$^2$, which suggests that
there is also a significant anisotropy in $\chi\prime\prime(q, \omega_n$).
$(1/T_1T)_c$ and $(1/T_1T)_{ab}$ are related to $\chi\prime\prime_{in}$
(in plane) and $\chi\prime\prime_{out}$ (out of plane) by the following
relations,\cite{Ishida01}
\begin {equation}
(1/T_1T)_c \propto \sum_q 2|H_q^{in}|^2
\frac{\chi\prime\prime_{in}(q, \omega)}{\omega_n},
\end {equation}
\begin {equation}
(1/T_1T)_{ab} \propto \sum_q [|H_q^{out}|^2
\frac{\chi\prime\prime_{out}(q,\omega)}{\omega_n} + |H_q^{in}|^2
\frac{\chi\prime\prime_{in}(q,\omega)}{\omega_n}],
\end {equation}
\begin{figure}
{\centering {\includegraphics{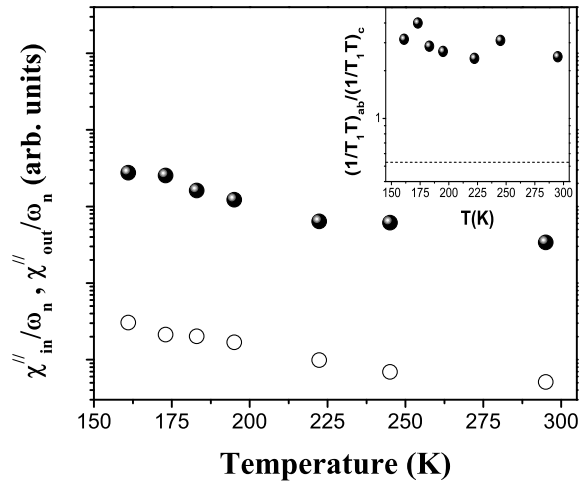}}} \caption{Variation of $\chi\prime\prime_{in}$/$\omega_n$ (open circle) and
$\chi\prime\prime_{out}$/$\omega_n$ (filled circle)with respect to T for
SmCoPO. Inset: (1/T$_1$T)$_{ab}$/(1/T$_1$T)$_c$ versus T and the
dashed line indicates the value of
[(H$_{hf}^{ab}$)$^2$ + (H$_{hf}^c$)$^2$]/2(H$_{hf}^{ab}$)$^2$ $\approx$ 0.53.}
\end{figure}
Using these relations, we have estimated
$\chi\prime\prime_{in}$/$\omega_n$ and
$\chi\prime\prime_{out}$/$\omega_n$, whose $T$ dependence are shown
in Fig. 6. $\chi\prime\prime_{out}$/$\omega_n$ is about two orders
of magnitude higher than $\chi\prime\prime_{in}$/$\omega_n$. Such
anisotropy in relaxation rate was also observed in other FeAs based
systems like BaFe$_2$As$_2$,\cite{Kitagawa08}
SrFe$_2$As$_2$,\cite{Kitagawa09} LaFeAsO,
LaFeAsO$_{1-x}$F$_x$\cite{Ishida10} and also Co doped
BaFe$_2$As$_2$\cite{Ning09,Ning10,Ishida10} systems. Define
$\textbf{\textit{A}}$  =
[(1/T$_1$T)$_{ab}$/(1/T$_1$T)$_c$]/[{(H$_{hf}^{ab}$)$^2$ +
(H$_{hf}^c$)$^2$}/2(H$_{hf}^{ab}$)$^2$]. $\textbf{\textit{A}}$ will
be one if their is no anisotropy in $\chi\prime\prime$/$\omega_n$
and $\textbf{\textit{A}}$ will deviate from unity if anisotropy comes from
$\chi\prime\prime$/$\omega_n$. The values of $\textbf{\textit{A}}$ obtained using the reported $^{75}$As NMR $T_1$ data in  BaFe$_2$As$_2$, SrFe$_2$As$_2$,
and LaFeAsO are 2.4, 1.75, 1.5 respectively. Also in Co doped BaFe$_2$As$_2$,
(1/T$_1$T)$_{ab}$/(1/T$_1$T)$_c$ is between 1-2. The value of $\textbf{\textit{A}}$ in
case of SmCoPO is 5.66. This clearly indicates a dominant contribution of Sm 4$f$ spin fluctuations over that due to Co-3$d$ electrons, in the
anisotropy of 1/$T_1$ in SmCoPO.
Anisotropy in spin-fluctuations arises from the spin-orbit coupling for
which the spin and orbital degrees of freedom get mixed and dynamic
magnetic susceptibility becomes anisotropic. So the preferred
directions of orbital fluctuations are determined by the geometry
and orbital characters of the Fermi surfaces. Band structure
calculations using tight-binding approximation is highly needed to
evaluate which 3$d$ and 4$f$ orbital fluctuations are responsible for
anisotropy in spin-fluctuations in SmCoPO.

When the Knight shift and nuclear spin-lattice relaxation process are governed by
conduction electrons, $1/T_1TK^2$ is constant. If their is an
exchange interaction between the electrons then using Stoner
approximation along with random phase approximation, modified Korringa
relation can be written as\cite {Moriya63,Narath68,Lue99}
$S_0/T_1TK_{spin}^2 = K(\alpha$), where $S_0 = (\hbar/4\pi
k_B)(\gamma_e/\gamma_n)^2$ and
\begin {equation}
K(\alpha) = \langle(1-\alpha_0)^2/(1-\alpha_q)^2\rangle_{FS}.
\end {equation}
$\alpha_q = \alpha_0\chi_0(q)/\chi(0)$ is the q-dependent
susceptibility enhancement, with $\alpha_0$ = 1 - $\chi_0(0)/\chi(0)$
representing the $q=0$ value. The symbol $\langle\rangle_{FS}$ means the average over q space on
the Fermi surface. $\chi$(0) and $\chi_0(q)$ represents
the static susceptibility and the $q$ mode of the generalized
susceptibility of noninteracting electrons respectively.
 $K(\alpha)<$ 1 means the spin-fluctuations are
enhanced around $q=0$, leading to the predominance of ferromagnetic
correlations and $K(\alpha) >$ 1 signifies that spin-fluctuations
are enhanced away from $q=0$. This would indicate a tendency towards
AF ordering (at $q\neq$0). Inset of Fig. 7. shows that for LaCoPO,
$K(\alpha)_{ab}$ and $K(\alpha)_c$ are $<$ 1 when calculated at 160
K, which shows the predominance of ferromagnetic spin-fluctuations
in both these directions. Whereas, in SmCoPO, $K(\alpha)_{ab}$ $<$ 1
but $K(\alpha)_{c}$ $>$ 1/(1.8), calculated at 160 K. The values of
$K(\alpha)_{ab}$, $K(\alpha)_{c}$ for LaCoPO and the value
$K(\alpha)_{ab}$ for SmCoPO suggest that the spin fluctuations are
ferromagnetic in nature both in the $ab$-plane and along the $c$
direction in LaCoPO. However, the same in SmCoPO is FM in the
$ab$-plane, while along $c$-direction there is  a signature of the
presence of $q\neq$0 modes in addition to $q=0$ modes. Thus in SmCoPO their exist weak AFM
spin-fluctuations along c axis in contrast to LaCoPO (inset of Fig.
7).

According to the theory of weak itinerant ferromagnet, if 3D/2D
spin-fluctuations are dominant\cite {Hatatani95, Ishigaki98} then $1/T_1T$ $\propto$
$\chi^{1(3/2)}$. Fig. 8 shows the
$T$ versus $(1/T_1TK)_{ab}$ and $(1/T_1TK^{3/2})_{ab}$ plots revealing
dominant 2D FM spin-fluctuations in the $ab$ plane of SmCoPO particularly in the paramagnetic
region.

\begin{figure}
{\centering {\includegraphics{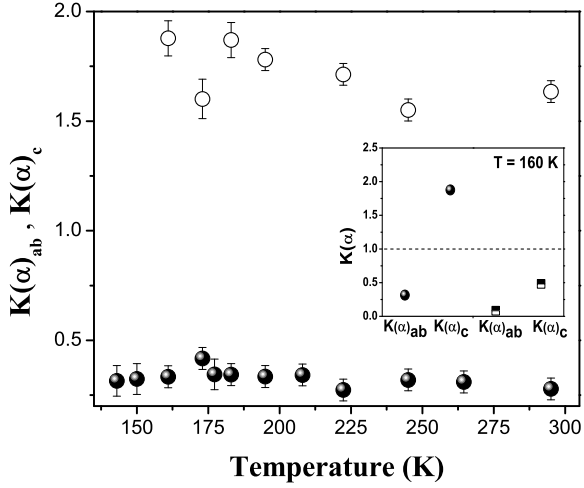}}} \caption{$K(\alpha)_{ab}$ (filled circle) and $K(\alpha)_c$ (open
circle) versus T for SmCoPO.
Inset shows K($\alpha$)$_{ab}$, K($\alpha$)$_c$ (filled circle) for
SmCoPO and K($\alpha$)$_{ab}$, K($\alpha$)$_c$ (half-filled square)
for LaCoPO at 160 K and the dashed line corresponds to K($\alpha$) = 1.}
\end{figure}

\subsubsection{Spin fluctuations parameters}
Following the theory of Ishigaki and Moriya\cite{Ishigaki98} one can write
the imaginary part of dynamic spin susceptibility in terms of the two
spin-fluctuation parameters  $T_0$ and $T _A$ which characterize the
width of the spin excitations spectrum in frequency and wave vector
(\textbf{q}) space respectively. For  ferromagnetic correlations, we have
\begin {equation}
\chi(\textbf{q},\omega) = \frac{\pi
T_0}{\alpha_QT_A}(\frac{x}{k_B2\pi T_0x(y+x^2)-i\omega\hbar}),
\end {equation}
where $x=\textbf{q}/q_B$ with $q_B$ being the effective zone
boundary vector, $\alpha_Q$ a dimensionless interaction constant
close to unity for a strongly correlated system, $y =
1/2\alpha_Qk_BT_A\chi(0,0)$. Here the susceptibility is per spin and
in units of 4$\mu_\mathrm{B}^2$ and has the dimension of inverse of
energy, $T_0$ and $T_\mathrm{A}$ are in Kelvin. From Eq. 11
 one
can derive $\chi\prime\prime(\textbf{q},\omega_n)$ in the limit
$\omega_n$ $\rightarrow$ 0, since $\hbar\omega_n$ $\ll$ $k_BT$. For
3D spin fluctuations governing the relaxation process, one has to
integrate $\chi\prime\prime(\textbf{q},\omega_n)/\omega_n$, over a
sphere of radius $\textbf{q}_B(\frac{6\pi^2}{v_0})^{1/3}$,
whereas, in case of 2D spin fluctuations, the integration has to be
done over a disc of radius
$\textbf{q}_B(\frac{4\pi}{v_0})^{1/2}$. $v_0$ corresponds to the atomic volume of Co. So in the latter case
\begin {equation}
1/T_1T = \gamma_n^2 H_{hf}^2 /4T_AT_0y^{3/2} + \alpha
\end {equation}
where according to 2D SCR theory, $y$ can be approximately written as $y
= (T/6T_0)^{2/3}\exp(-p^2T_A/10T)$, where $p$ is the ferromagnetic moment in $\mu_B$ units
 and $\alpha$ is the temperature independent
contribution of $1/T_1T$ arising from the orbital moment of $p$ and
$d$ electrons and the spin of the conduction electrons. Using Eq.
16, we have estimated the spin-fluctuation parameters $T_A$ and
$T_0$ in SmCoPO as 21000 and 2043 K respectively. The dashed line of
Fig. 5 corresponds to Eq. 16.

\subsubsection{Spin fluctuations and possible antiferromagnetic spin-structure}
According to the SCR theory of itinerant antiferromagnetism if
3D(2D) spin-fluctuations governs the relaxation
process\cite{Ishigaki98} then $1/T_1T$ is proportional to
$\chi^{1/2,(1)}$. Inset of Fig. 8 shows that ($1/T_1T)_c$ is nearly
proportional to the intrinsic spin susceptibility, probed by NMR
shift $K_c$ (which is proportional to $\chi_c$), which reveals that
AFM spin-fluctuations are also nearly 2D in nature. A small slope in
this plot could be a signature of the presence of weak FM spin
fluctuations superimposed on the weak 2D AFM one along the
c-direction. Presence of FM spin fluctuation along c-direction due
to the inter layer exchange interaction, in the paramagnetic phase
in SmCoPO is relevant (observed from $^{139}$La NMR in
LaCoPO as shown in \cite{Majumder10}) as the FM transition precedes the AFM
transition. Possibly this is the reason for which
$\chi\prime\prime_{in}$/$\omega_n$ $<$
$\chi\prime\prime_{out}$/$\omega_n$, because
$\chi\prime\prime_{out}$/$\omega_n$ arises from the sum of the
contributions from $q$ = 0 and $q$ $\neq$ 0 mode of
spin-fluctuations. The appearance of weak AFM spin-fluctuations of the Co-3$d$ spins,
along $c$-direction, far above the ferromagnetic transition
(which was absent in case of LaCoPO) is a signature of AF exchange interaction between the Sm-4$f$ and the Co-3$d$ electrons. This is consistent with the fact that Sm-O plane is situated in between two Co-P planes and both are parallel to $ab$ plane. Due to the presence of AF interaction along $c$-direction, though the Co spins in each Co-P plane order
ferromagnetically below $T_C$, at lower temperature when the
AF exchange interaction between the Sm-4$f$ and Co-3$d$ spins, along the $c$-direction becomes
more stronger, the Co-3$d$ spins in the two adjacent planes would
try to align antiparallel even if they remain parallel to each other
within a plane. As  a result, the system orders
antiferromagnetically. Such type of spin structure in NdCoAsO has
recently been proposed from elastic neutron scattering
study.\cite{McGuire10}
Signature of the presence of weak AF exchange interaction between the Sm-4$f$ and Co-3$d$ spins even above $T_C$ obtained from $^{31}$P spin-lattice relaxation data, could
possibly be the reason for the persistence of AF transition in SmCoPO even in a magnetic field of 14 T, whereas, it disappears
completely at $H$=5 T in NdCoPO, though the Nd 4$f$ effective moment
is higher than that of Sm 4$f$. As the lattice parameter $c$ in
SmCoPO is less than that in NdCoPO, therefore, the strength of the AF exchange interaction between Sm-4$f$ and Co-3$d$ spins
present along the $c$-direction could be more stronger in SmCoPO than in NdCoPO. Therefore, the $^{31}$P NMR shift and the spin-lattice relaxation studies in other members of \emph{L}CoPO series would be
interesting to understand the effect of the change in the lattice
volume, on the nature of the spin fluctuations and as well as the
extent of anisotropy, which drives the system to a particular ground
state.

\begin{figure}
{\centering {\includegraphics{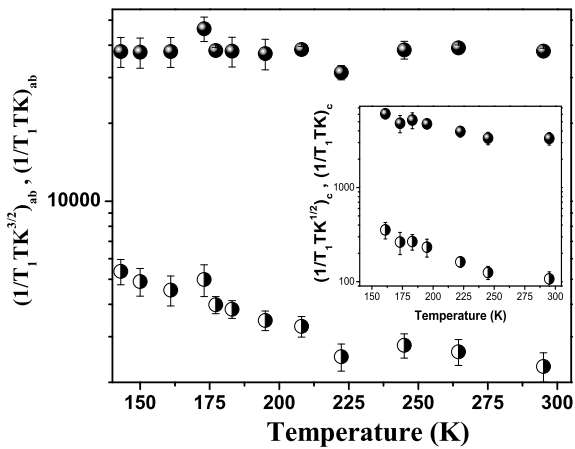}}} \caption{(1/T$_1$TK$^{3/2}$)$_{ab}$ (filled circle),
(1/T$_1$TK)$_{ab}$ (half-filled circle) versus T for SmCoPO and
inset shows (1/T$_1$TK$^{1/2}$)$_c$ (half-filled circle),
(1/T$_1$TK)$_c$ (filled circle) versus T for SmCoPO}
\label{structure}
\end{figure}

\section{CONCLUSIONS}
We have reported $^{31}$P NMR results in the powder sample of SmCoPO.
 The spectral features reveal an axially symmetric nature of the local magnetic field.
 At low temperature, the anisotropy of the internal magnetic field increases, with $K_{ab}$
 increasing faster than that of $K_c$. The intrinsic width 2$\beta$ shows a linear variation with $\chi_M$ in the range 300 - 170 K. Deviation from linearity below 170 K arises due to the enhancement of ($1/T_2)_{dynamic}$.
 This enhancement of ($1/T_2)_{dynamic}$ along with the continuous increase of anisotropy in the internal magnetic field is responsible for the wipe out of the NMR signal, well above $T_C$. Absence of anisotropy in $1/T_2$ indicates the isotropic nature of the longitudinal component of the fluctuating local magnetic field.

  Observed large anisotropy in $1/T_1$ of SmCoPO  compared to that of LaCoPO, confirms a significant contribution of Sm-4$f$ electron arising from indirect RKKY interaction. This indicates a non-negligible hybridization between Sm-4$f$ orbitals and the conduction band, over and above the itinerant character of the Co-3$d$ spins. The anisotropy of $1/T_1$ originates mainly from the orientation dependence of $\chi^{\prime\prime}$($\textbf{q}$,$\omega$). The 3$d$-spin fluctuations in the $ab$ plane is primarily of 2D FM in nature similar to that in LaCoPO, while along the $c$-axis, a signature of weak 2D AFM spin-fluctuations superimposed on weak FM spin-fluctuations even in a field of 7 T and far above $T_C$ is observed. The enhancement of this AFM exchange interaction below $T_C$ should be responsible to drive the Co moment to an AFM ordered state.

\end{document}